\documentclass{article}
\usepackage[dvips]{graphicx}
\usepackage{epsfig}
\usepackage{amssymb}
\usepackage{multicol}
\newcommand{\be}{\begin{equation}}
\newcommand{\ee}{\end{equation}}
\newcommand{\ecoli}{{\it E. coli}}
\newcommand{\aureus}{{\it S. aureus}}
\newcommand{\yeast}{{\it S. cerevisiae}}

\begin{document}


\noindent
{\LARGE \bf Low Degree Metabolites Explain Essential Reactions
and Enhance Modularity in Biological Networks}

\vspace{0.4cm}
\noindent
\renewcommand{\thefootnote}{\fnsymbol{footnote}}
{\small \bf Areejit Samal$^1$, Shalini Singh$^1$, Varun Giri$^1$,
Sandeep Krishna$^{2}$ \footnote[2]{Present address: Niels Bohr
Institute for Astronomy, Physics and Geophysics, Blegdamsvej 17,
Copenhagen DK-2100, Denmark}, N. Raghuram$^3$ $\&$ Sanjay
Jain$^{1,4,5}$ \footnote[1]{Corresponding author
(jain@physics.du.ac.in)} }


\vspace{0.3cm}

\begin{center}
{\small \it
$^1$Department of Physics and Astrophysics, University of Delhi,
Delhi 110007, India \\
$^2$National Centre for Biological Sciences, UAS-GKVK Campus, Bangalore
560065, India \\
$^3$School of Biotechnology, GGS Indraprastha University, Delhi
110006, India \\
$^4$Jawaharlal Nehru Centre for Advanced Scientific Research, Bangalore 560064,
India \\
$^5$Santa Fe Institute, 1399 Hyde Park Road, Santa Fe, NM 87501, USA\\
}
\end{center}


\vspace{1cm}
\begin{center}
{\bf \large Abstract}
\end{center}
\vspace{0.4cm}

\noindent
{{\bf Background:} \\}
\noindent
Recently there has been a lot of interest in identifying modules at the level
of genetic and metabolic networks of organisms, as well as in identifying
single genes and reactions that are essential for the organism.
A goal of computational and systems biology is to go beyond identification
towards an explanation of specific modules and essential genes and reactions
in terms of specific structural or evolutionary constraints.\\
\noindent {{\bf Methodology:} \\} \noindent In
the metabolic networks of {\it Escherichia coli}, {\it
Saccharomyces cerevisiae} and {\it Staphylococcus aureus}, we
identified metabolites with a low degree of connectivity,
particularly those that are produced and/or consumed in just a
single reaction. Using flux balance analysis (FBA) we also
determined reactions essential for growth in these metabolic
networks. We find that most reactions identified as essential in
these networks turn out to be those involving the production or
consumption of low degree metabolites. Applying graph theoretic
methods to these metabolic networks, we identified connected
clusters of these low degree metabolites. The genes involved in
several operons in \ecoli\ are correctly predicted as those of
enzymes catalyzing the reactions of these clusters. Furthermore,
we find that larger sized clusters are over-represented in the
real network and are analogous to a `network motif'. Using FBA for
the above mentioned three organisms we independently identified
clusters of reactions whose fluxes are perfectly correlated. We
find that the composition of the latter `functional clusters' is
also largely explained in terms of
clusters of low degree metabolites in each of these organisms. \\
{{\bf Conclusion:} \\} \noindent Our
findings mean that most metabolic reactions that are essential can
be tagged by one or more low degree metabolites. Those reactions
are essential because they are the only ways of producing or
consuming their respective tagged metabolites. Furthermore,
reactions whose fluxes are strongly correlated can be thought of
as `glued together' by these low degree metabolites. The methods
developed here could be used in predicting essential reactions and
metabolic modules in other organisms from the list of metabolic
reactions. \\


\newpage

\section{Introduction}
\noindent
Evolution has produced organisms that are robust to
various perturbations, yet the specific knockout of a single gene
can be lethal to the organism. Similarly, organisms have some
redundancy in their metabolic pathways, but single reactions whose
knockout brings the growth of a cell to a halt
--- called `essential' reactions --- are also known to exist in metabolic
networks \cite{EP2000,DHP2004,CKRHP2004}. What properties of a
specific gene or reaction, within the context of the overall
structure and organization of biochemical networks, make it
essential for the organism? We show that most essential metabolic
reactions in {\it Escherichia coli} \cite{RVSP2003}, {\it
Saccharomyces cerevisiae} \cite{DHP2004}\ and {\it Staphylococcus
aureus} \cite{BP2005}\ can be explained by the fact that they are
associated with a low degree metabolite. Metabolic and protein
interaction networks contain nodes with a large variation in their
degree of connectivity \cite{JTAOB2000,WF2001,JMBO2001}. In case
of protein interaction networks it has been suggested that
essentiality of a protein is correlated with its degree
\cite{JMBO2001}. Hence, protein interaction networks are
vulnerable to removal of highly connected proteins called `hubs'.
In contrast, for metabolic networks, one is usually interested in
the essentiality of reactions rather than metabolites. Recently,
Mahadevan and Palsson \cite{MP2005}\ have shown that low degree
metabolites are almost as likely to be associated with essential
reactions as high degree metabolites. We show here that in fact
{\it almost all} essential reactions are explained by virtue of
being tagged to some {\it low} degree metabolite.

Another theme in systems and computational biology has been to
identify genetic regulatory modules
\cite{ESBB1998,IFBSZB2002,SSRPBKF2003}, functional clusters
\cite{PSNMS1999}-\cite{RP2004}\ and graph-theoretic modules
\cite{GJC2003,GA2005}\ in metabolic networks. Modularity of
complex biological networks contributes to the robustness,
flexibility, and evolvability of organisms, and also towards
making their organization more comprehensible \cite{HHLM1999}.
What structural features of metabolic networks cause specific
subsets of metabolic reactions to have strongly correlated fluxes?
We observe that low degree metabolites lead to one such structure
in the metabolic network. Such metabolites contribute to a
rigidity or coherence of reaction fluxes in the network resulting
in clusters of highly correlated reactions. For example, in any
steady state, where the concentrations of all metabolites are
constant, a metabolite that can be produced in only one reaction
and consumed in only one causes both reactions to have equal (or
proportional with a fixed proportionality constant) fluxes.
Maintaining the metabolic network close to a steady state then
requires enzymes for both reactions to be simultaneously active,
and hence the corresponding genes to be co-expressed, resulting in
a transcription module containing those genes. In this work we
first locate metabolites based purely on their low degree in the
metabolic network. Then we show that clusters of their reactions
predict genetic regulatory modules, as captured in the structure
of operons \cite{regulondb,ecocyc}, with a high probability in
\ecoli. Furthermore, the composition of most functional clusters
is also explained via the low degree clusters embedded inside
them.

Biological networks have two properties that are currently
regarded as unrelated: One, they have functional modules, and two,
they have single genes or metabolic reactions whose knockout is
lethal. An implication of the present work is that in metabolic
networks, both properties can arise as consequences of the same
structural property: the existence of low degree metabolites. Our
work provides an {\it explanation}, rather than just
identification, of essential reactions and metabolic modules.


\section{Lowest degree metabolites and their clusters.}

\noindent

A metabolite may be designated as `uniquely produced' or `UP'
(`uniquely consumed' or `UC') if, in the bipartite graph of
reactions and metabolites, the node corresponding to the
metabolite has in-degree (out-degree) equal to unity; in other
words, if there is only one reaction in the network that produces
(consumes) the metabolite. A metabolite that is both UP and UC (a
`UP-UC metabolite') has the lowest degree in the network. Examples
of UP-UC metabolites taken from the metabolic networks
\cite{RVSP2003,BP2005}\ of \ecoli\ and \aureus\ can be seen in
Fig. 1. In any metabolic steady state the concentration of a
metabolite is fixed; its rate of production is equal to that of
consumption. Hence for a UP-UC metabolite in any steady state, the
flux of the reaction producing it is proportional to that of the
reaction consuming it, with the proportionality constant
determined by the stoichiometric coefficients of the metabolite in
the two reactions. A `UP-UC cluster' of reactions may be defined
as a set of reactions connected by UP-UC metabolites. In a steady
state fixing the flux of any reaction in the UP-UC cluster fixes
the fluxes of all other reactions in the cluster (see Fig. 1).
These clusters include linear pathways but can also have branched
or cyclic structure. UP-UC clusters are special cases of
reaction/enzyme subsets \cite{PSNMS1999,SKWMP2002,SKBSG2002} and
fully coupled reactions or co-sets \cite{PPP2002,BNSM2004,RP2004}.
Each UP-UC cluster of reactions can be replaced by an effective
reaction without affecting the steady state performance and can be
used to coarse-grain metabolic networks
\cite{PSNMS1999,SKWMP2002}.

A reaction was designated as `uniquely producing' or `UP'
(`uniquely consuming' or `UC') if it produced (consumed) a UP (UC)
metabolite. The number of UP (UC) reactions in the metabolic
networks of \ecoli, \yeast\ and \aureus\ were found to be 289
(272), 391 (370) and 277 (218), respectively, while the number of
reactions that are either UP or UC or both (we refer to this set
as `UP/UC reactions') is 417, 583 and 376. We will show below that
such reactions play a special role in metabolic networks.


\section{Results}


\subsection{Essential reactions are largely explained by UP/UC structure}

\noindent We used the flux balance analysis (FBA)
\cite{EP2000,VP1994,EIP2001,SVC2002}\ approach to determine
essential reactions in the metabolic networks of \ecoli, \yeast\
and \aureus. We computed the steady state optimal flux vectors for
each of these organisms in aerobic conditions for all permissible
single organic carbon sources in a minimal medium. We found a
feasible solution (with a nonzero growth rate) for 89, 43 and 27
sources in \ecoli, \yeast\ and \aureus\ respectively. The list of
feasible carbon sources under minimal media in these organisms is
provided in Supplementary Tables S1, S2 and S3.

We considered the effect of `switching off' reactions (by setting
their maximum flux equal to zero) one by one, on the optimal
growth rate for each food source. A reaction was designated as
`essential' for a particular food source if switching it off
resulted in a zero optimal growth rate under that input condition.
We designated a reaction as `globally essential' for an organism
if it was essential for all its feasible minimal media under
aerobic conditions. The number of essential reactions for each
minimal media varied between 200 and 240 reactions and the number
of globally essential reactions was 164 for the \ecoli\ metabolic
network. Similarly, we found that the number of globally essential
reactions in metabolic networks of \yeast\ and \aureus\ were 127
and 196 respectively.

\subsubsection{Most essential reactions either produce or consume
a UP or UC metabolite}

\noindent
Of the 164 globally essential reactions in the \ecoli\
metabolic network, 133 were found to be either UP or UC.
Similarly, we found a high fraction of globally essential
reactions in metabolic networks of \yeast\ and \aureus\ to be UP
or UC (see Table 1). This explains why this subset is essential:
there is simply no other path around these reactions in the entire
network to produce or consume some metabolite that is presumably
required for the eventual production of biomass. In a recent paper
\cite{MP2005}\ Mahadevan and Palsson have determined, for each
metabolite in the network, the fraction of its reactions that are
essential. They have observed that this `lethality fraction' of
the low degree metabolites is on average comparable to high degree
metabolites, and in particular, some metabolites with in and out
degree unity (that we have designated here as UP-UC metabolites)
have lethality fraction unity. We present here a stronger result
regarding the role of low degree metabolites: most essential
reactions involve at least one UP or UC metabolite. These
reactions may involve other metabolites of higher degree, but
their essentiality is due to their uniqueness in producing or
consuming a low degree metabolite.

\subsubsection{The correspondence between essential and UP/UC reactions
is even tighter in the `reduced network'}

\noindent To understand the remaining globally essential
reactions, we considered a reduced or pruned version of the
network. Certain reactions in
various reconstructed metabolic networks are such that they have a zero
flux value under all steady states for stoichiometric reasons.
These reactions are referred to as `strictly detailed balanced' reactions
\cite{SS1991} or `blocked' reactions \cite{BNSM2004}, and can
be removed from the network for any steady state analysis.
We used a previously described algorithm
\cite{BNSM2004}\ to determine blocked reactions in the metabolic
networks of \ecoli, \yeast\ and \aureus. We found 290 (800, 294)
of the 1176 (1579, 865) reactions in the \ecoli\ (\yeast, \aureus)
metabolic network to be blocked. We removed the blocked reactions
from each network to obtain the `reduced network' for each
organism (containing 886, 779 and 571 reactions respectively).

Note that the essential reactions obtained by implementing FBA on
the reduced network are exactly the same as those obtained from
the original network for each input condition. Hence, instead of
requiring a metabolite to be UP or UC across the entire metabolic
network, we asked if it was UP or UC in the reduced network. The
set of UP(UC) metabolites and reactions so obtained turns out to
be somewhat smaller than the original set. In \ecoli, \yeast\ and
\aureus\ the new set of UP/UC reactions has 352, 306 and 276
reactions. This is so because several reactions that were UP/UC in
the original network happen to be blocked and are now removed.
Conversely some metabolite that was earlier not UP(UC) can now
become UP(UC) after the removal of a reaction. This adds new
reactions to the UP/UC set but this number turns out to be smaller
than the number removed (details are given in Supplementary Table
S4). The new UP(UC) metabolites have, by definition, their in
(out) degree unity in the reduced network; even in the original
network they have a low degree (for \ecoli\ their average in (out)
degree in the original network is 1.31 (1.33)). We emphasize that
the reduced network as defined above and hence the set of new
UP(UC) reactions is uniquely determined by the original network.

We found that 156 out of the 164 globally essential reactions (95
$\%$) in the \ecoli\ metabolic network to be UP or UC in the
reduced network. Similarly, we found that almost all globally
essential reactions in \yeast\ and \aureus\ were either UP or UC
in the reduced network (92 and 93 $\%$ respectively; see Table 1)
thereby underscoring the fact that nodes with a low degree of
connectivity play an `essential' role in metabolism. The
importance of low-degree nodes in the essential functionality of
complex autocatalytic networks has also been observed elsewhere
\cite{JK2002} in a different context.

This finding provides some insight into the structural or
topological origin of essential reactions in metabolic networks.
It is, of course, obvious that if a certain metabolite is an
essential intermediate for the production of some biomass
metabolite, and if this metabolite is uniquely produced or
uniquely consumed, then the corresponding production or
consumption reaction will be essential for the growth of the cell.
However the converse of this statement --- that all essential
reactions in the network should have this topological property ---
is far from obvious. Our finding that about 5-8 $\%$ of essential
reactions do not have this property proves that the converse
statement is indeed false. Thus the fact that the overwhelming
majority (92-95 $\%$) of essential reactions have this topological
property is a characterization of the nature of metabolic networks
found in organisms. We remark that we do not as yet understand why
the remaining essential reactions happen to be essential.

\subsubsection{Most UP/UC reactions are essential in
some condition or other}

\noindent We found that there are 352 UP or UC metabolic reactions
in the \ecoli\ reduced network. 156 of these 352 reactions were
globally essential, while 288 of these 352 reactions (82 \%) were
essential for at least one of the 89 possible minimal media in
\ecoli. Some of these UP/UC reactions were part of the input
pathways of only one carbon source, hence they were essential only
for that input. In \yeast\ 170 out of 306 UP/UC reactions (56 \%)
in the reduced network are essential in at least one input
condition, while in \aureus\ 257 out of 276 (93 \%) have this
property. The substantial difference between \yeast, a eukaryote,
and the two bacteria may reflect a more evolved metabolic
structure that needs to be further investigated.

\subsubsection{Comparison between computationally determined essential
reactions and lethal single gene knockouts}

\noindent To check the agreement of essential reactions in the
\ecoli\ metabolic network with a database \cite{Gerdes2003}\ of
experimentally determined essential genes in a rich medium, we
implemented FBA for a rich medium containing all food sources for
the \ecoli\ metabolic network \cite{AKVOB2004}. We found 95
reactions to be essential in this medium for \ecoli. 89 of these
95 reactions were found to be either UP or UC in the reduced
network. Of the 95 essential reactions in rich medium, information
about the corresponding genes was available for only 85 reactions.
Of these 85, 14 reactions had known isozymes, i.e, multiple
enzymes catalyzing the reaction, hence the corresponding genes are
not expected to be essential. Of the remaining 71 reactions, 5 had
associated genes whose essentiality was undetermined in the
database. Of the remaining 66 reactions, 38 reactions had
associated genes that had been found to be essential in the
database \cite{Gerdes2003}, which is a fairly high fraction.
Conversely, of the 618 essential genes determined for {\it E.
coli} by Gerdes et al, 158 genes were also part of the \ecoli\
metabolic network \cite{RVSP2003}\ used for our study. 103 of the
above 158 essential genes had their products catalyzing only a
single reaction in the \ecoli\ metabolic network. Of these 103
essential genes, 62 were associated with a UP or UC reaction.
Further, using the reduced network, we found that 73 of the 103
essential genes were associated with a UP or UC reaction. The
discrepancy between theoretical prediction and experimental data
may be reconciled by the incomplete knowledge about possible
isozymes for certain reactions or uncharacterized alternative
metabolic pathways in the present in-silico metabolic model
\cite{CKRHP2004}.


\subsection{Low degree clusters predict regulatory modules}

\noindent We found that the \ecoli\ metabolic network
\cite{RVSP2003} contained 185 UP-UC metabolites. We determined all
UP-UC clusters in the network (see methods). The total number of
UP-UC clusters in \ecoli\ metabolic network was found to be 85;
their size distribution is shown by the grey bars in Fig. 2. The
list of all reactions in each UP-UC cluster for the \ecoli\
metabolic network is given in Supplementary Table S6. We then
investigated whether the genes coding for the enzymes of the
reactions in a UP-UC cluster are part of the same operon in
\ecoli. Genes on the same operon are by definition part of a
genetic module since they are coregulated. At the moment genes
corresponding to enzymes of reactions of the network have been
identified for only part of the network. Of the 85 UP-UC clusters
in the \ecoli\ metabolic network, only 69 clusters had two or more
reactions with known corresponding genes. We looked at the
regulation of these 69 UP-UC clusters using the known operon
information from RegulonDB \cite{regulondb} and Ecocyc
\cite{ecocyc} databases. Genes (of reactions within UP-UC
clusters) that belong to the same operon in \ecoli\ are indicated
in Supplementary Table S5. For 42 of the 69 UP-UC clusters, we
found that two or more genes of the cluster were part of the same
operon. Further, 36 of these 42 UP-UC clusters had at least half
of their genes belonging to the same operon. We also found that 21
UP-UC clusters have at least one possible set of constituent genes
catalyzing all reactions in the cluster belonging to the same
operon.

To show that two genes belonging to a UP-UC cluster in \ecoli\
have greater probability of lying on the same operon than
otherwise expected, we performed the following test. We found 251
unique genes catalyzing various reactions in the 69 UP-UC
clusters. If we randomly pick any two of these 251 genes, the
probability that the two genes lie on the same operon is 0.0057.
If we randomly pick a pair of genes that belong to the same UP-UC
cluster from this set of 251 genes, the probability that the two
genes lie on the same operon is 0.29. Thus regulatory modules are
predicted correctly with a high probability by this method. It is
possible that UP-UC clusters will find even greater correspondence
with regulatory modules when expression data is analysed; our
comparison rests only on operon data, and only about 25 percent of
the transcriptional regulatory network of \ecoli\ is presently
believed to have been identified \cite{CKRHP2004}. It would also
be interesting to extend this analysis to the other two organisms.



\subsection{Large UP-UC clusters are analogous to network motifs}

\noindent We asked the question: Is it expected that a network
like the \ecoli\ metabolic network of 618 metabolites and 1176
reactions with 185 UP-UC metabolites will have a distribution of
UP-UC clusters as given in Figure 2? To answer this question, we
compared the distribution of UP-UC clusters in the real \ecoli\
metabolic network with a suitably randomized version of the
network \cite{SMMA2002}. The randomized network has the same
number of metabolite nodes and reaction nodes and the same number
of incoming and outgoing links at each node as the real \ecoli\
metabolic network (see methods). Averaging over 1000 realizations
of the randomized metabolic network we found a cluster
distribution as shown by the black line in Fig. 2.

This shows that the actual metabolic network of \ecoli\ has its
UP-UC metabolites bunched up next to each other, forming larger
clusters than expected in random networks with the same local
connectivity properties. Thus, larger size (size $\geq 8$) UP-UC
clusters are over-represented in the real \ecoli\ metabolic
network, and may be collectively considered as analogous to a
network {\it motif} (as defined in \cite{SMMA2002,MSIKCA2002}),
while smaller size ($\leq 3$) UP-UC clusters are under-represented
in the real network, and may be collectively considered analogous
to an `anti-motif' \cite{MIKLSASA2004}. We also found
qualitatively similar results for the metabolic networks of
\yeast\ and \aureus\ (data not shown).



\subsection{Low degree metabolites explain perfect clusters}

\noindent Correlated reaction sets are sets of reactions in the
metabolic network that are always used together in functional
states of the network. Each flux vector obtained using FBA
represents one possible functional state of the network. For each
feasible minimal medium we obtained one flux vector with a nonzero
growth rate. We defined an `active' reaction as one that had a
nonzero flux in at least one of the latter flux vectors. Then we
computed the correlation coefficient among fluxes of the active
reactions across these flux vectors in a manner analogous to the
correlation of gene activity from microarray data across different
conditions \cite{ESBB1998}\ (see methods). A `perfect cluster' is
a set of reactions whose pairwise correlation coefficients with
each other are all unity across all sets of conditions. Reactions
in perfect clusters have fluxes that are proportional to each
other with the same proportionality constant under all the flux
vectors considered. We found that in the \ecoli\  metabolic
network, most of the 582 active reactions under 89 input
conditions were contained in several perfect clusters of size 2 or
more (see Table 2). These clusters, reported earlier in
\cite{TALK}\ overlap highly with the clusters of \cite{RP2004}.
One might ask: Why are particular subsets of reactions perfectly
clustered to each other. UP-UC clusters provide a structural
explanation for these perfect clusters. Of the 85 UP-UC clusters
in the entire \ecoli\ network, 46 UP-UC clusters are in the set of
active reactions. All the 46 active UP-UC clusters are subsets of
perfect clusters. To further explain the observed clustering of
reactions in the \ecoli\ metabolic network, we considered UP(UC)
metabolites in the reduced network. We found 94 UP-UC clusters in
the reduced network for \ecoli. Table 2 shows that most of the
perfect clusters in \ecoli\ are explained in terms of UP-UC
clusters in the reduced network in the sense that UP-UC clusters
account for the bulk of reactions in the perfect clusters. Most of
the co-sets reported in \cite{RP2004}\ for \ecoli\ are also
explained by UP-UC clusters in the reduced network (see
Supplementary Table S7). Further, we found that most perfect
clusters in the metabolic networks of \yeast\ and \aureus\ are
also explained by UP-UC clusters in their respective reduced
networks (see Supplementary Tables S8 and S9).


\section{Discussion}

\noindent In this paper we have observed that the lowest degree
metabolites are implicated in two distinct properties of the
metabolic networks, one, the existence of essential metabolic
reactions (and lethal single metabolic gene knockouts), and two,
existence of functional clusters in the metabolic networks (and
associated regulatory modules).

To some extent the identification of UP/UC metabolites depends on
the way the metabolic network is curated. For example, the
networks we have used leave out certain non-enzymatic reactions
such as protonation-deprotonation reactions. Since their inclusion
would render some of the presently UP(UC) metabolites non-UP(UC),
our definition of UP(UC) could be criticized as being somewhat
arbitrary. In this context it is worth noting that for the
networks as they stand, our definition of UP(UC) allows us to
establish a connection between distinct properties of the network
(e.g., between essentiality, a functional property and the UP/UC
character, a topological property), and that our main findings
hold for metabolic networks of three distinct organisms. This
suggests that UP/UC reactions as defined by us do capture a
certain pattern. In our view the important point is not that other
definitions of the network would obscure the pattern, but rather,
that there do exist systematic definitions of the network in which
a pattern is visible.

In metabolic networks the very existence of essential reactions is
an indicator of the fragility of the system: Even though the
network has many reaction nodes, the removal of a single essential
reaction node destroys the functionality of the network completely
by blocking the flow of an essential intermediate. Isozymes are a
way of dealing with this fragility. However, not all essential
reactions have isozymes \cite{PPH2004}; this means that evolution
has tolerated this fragility. Our finding that essential reactions
are tagged by low degree metabolites may provide some insight into
why this is the case. Metabolites that participate in very few
reactions perhaps do so in part because some feature of their
chemical structure prohibits ready association with other
molecules, i.e., their low degree is a consequence of constraints
coming from chemistry. Then evolution tolerates the reactions that
produce or consume such metabolites as essential because chemistry
leaves it no choice.

Alternatively, it could be that this fragility happens to be a
byproduct of some other desirable structural property that
contributes to robustness or evolvability, such as modularity. We
have drawn attention to the fact that low degree metabolites also
play a role in functional clustering of reactions in the metabolic
network. We have further provided evidence that the UP-UC clusters
at the metabolic level correspond, with a high probability, to
sets of genes forming modules at the regulatory level in \ecoli.

This raises the question: if low degree metabolites contribute to
modularity, could it be that the evolutionary advantages of that
have outweighed the disadvantage of the above mentioned fragility
caused by the same low degree metabolites? Is it the case that
evolution has preferred `chemically constrained' low degree
metabolites in spite of the fragility they cause because they
contribute to modularity? A goal in biology is to understand
highly evolved biological organization in terms of simpler and
more inevitable structures \cite{MOROWITZ1999}. Here we have
presented evidence that certain genetic regulatory modules, in
particular certain operons, mirror the low degree structure of the
metabolites whose production and consumption they regulate. This
could be an example of how the origin of certain regulatory
structure can be traced to simple chemical constraints.



\section{Methods}

\subsection{Detection of UP-UC clusters}

\noindent We used recently reconstructed metabolic networks of
\ecoli\ (version iJR904 \cite{RVSP2003}), \yeast\ (version iND750
\cite{DHP2004}) and \aureus\ (version iSB619 \cite{BP2005}) in
this study. The networks were downloaded from the site {\it
http:// gcrg.ucsd.edu/organisms/index.html}. Each reversible
reaction in the network was converted into two one sided
reactions. We excluded the external metabolites in the three
metabolic networks while determining the UP-UC metabolites. For
calculating various UP-UC clusters, we first identify all UP-UC
metabolites in the bipartite graph of the network. We then delete
all links in the graph except those going into and out of UP-UC
metabolites. From this new bipartite graph, we generate a
reaction-reaction graph, in which two reactions are connected if
one consumes a metabolite produced by the other. The weak
components of size $\geq 2$  of the reaction-reaction graph are
the various UP-UC clusters in the network.


\subsection{Generation of randomized networks}
\noindent We constructed the matrix $A = (A_{i\alpha})$ where
$A_{i\alpha}$ equals 1 if metabolite $i$ is produced in reaction
$\alpha$, -1 if it is consumed in reaction $\alpha$ and 0 if it
does not participate in reaction $\alpha$. Each nonzero entry of
$A$ defines a link in the bipartite graph of metabolites and
reactions. Starting from $A$ for the real network, we generated
randomized networks keeping the degree of each metabolite and
reaction node unchanged \cite{KTV1999,MS2002}. It is important to
distinguish between two kinds of links; one coming into a
metabolite node from a reaction node and the other going out of a
metabolite node to a reaction node. All the links or edges in this
bipartite graph were divided into these two groups. Two links are
then randomly selected in one of these two groups and swapped.
Before swapping, we ensure that the metabolite involved in any
link is not already involved in the reaction corresponding to the
other link. This process of selecting a random pair of links was
repeated 18000 times. We verified that more than 99.9\% of the
links were visited at least once. Starting from the real metabolic
network, this procedure is repeated 1000 times (with different
random number seeds), the UP-UC clusters determined for each of
the 1000 realizations of the randomized network and the average
taken thereof.


\subsection{Perfect clusters}
\noindent Using FBA we obtained $v^I_{\alpha}$, the velocity of
reaction $\alpha$ in an optimal steady state corresponding to
input condition $I$, $I = 1,\ldots,M$, $\alpha = 1,\ldots,N$,
where $M$ is the number of feasible minimal media and $N$ is the
number of distinct one way reactions in the metabolic network.
These $v^I_{\alpha}$ define the $M$ flux vectors we consider.
Given a set of $M$ flux vectors, the correlation coefficient
\cite{ESBB1998}\ between two active reactions $\alpha$ and $\beta$
is given by
$$C_{\alpha\beta} =
(1/M)\sum_{I=1}^M
v^I_{\alpha}v^I_{\beta}/(\phi_{\alpha}\phi_{\beta}),$$ where
$\phi_{\alpha} = [(1/M)\sum_{I=1}^M (v^I_{\alpha})^2]^{1/2}$.
Reactions $\alpha$ and $\beta$ are said to be perfectly correlated
in the given set of flux vectors if $C_{\alpha \beta}=1$ for that
set and all its subsets of flux vectors. A numerical value of
$C_{\alpha \beta} \geq 0.999999$ was taken as `unity' for this
purpose. Perfect clusters were identified by locating maximal sets
of reactions that were perfectly correlated to each other
pairwise.


\vspace{0.5cm}
\begin{center}
{\bf Acknowledgements\\}
\end{center}
\noindent \small {We thank Devapriya Choudhury, Shobhit Mahajan 
and Amitabha Mukherjee for discussions and helpful suggestions. 
AVADIS software was used for the initial visualization of 
clusters. We thank IUCAA
Reference Centre, University of Delhi for computational
infrastructure. We also thank S.N. Bose National Centre for Basic
Sciences, Kolkata and Centre for High Energy Physics, Indian
Institute of Science, Bangalore for infrastructure and hospitality
where part of this work was done. A.S. and S.S. acknowledge a
Senior Research Fellowship from CSIR and UGC, respectively.}




\newpage

\noindent
{\bf TABLES\\}
\vspace{0.4cm}

\hspace{1cm}
\begin{tabular}{|l|c|c|c|}
\hline
{\bf Organism} & {\bf \it E. coli } & {\bf \it S. cerevisiae } & {\bf \it S. aureus } \\
\hline
{\small Total number of reactions} & {\small 1176} & {\small 1579} & {\small 865} \\
\hline
{\small Number of globally essential reactions} & {\small 164 } & {\small 127 } & {\small 196 } \\
\hline
{\small Number of globally essential reactions } & & & \\
{\small that are UP or UC in the entire network} & {\small 133} & {\small 86 } & {\small 157 } \\
\hline
{\small Number of globally essential reactions } & & & \\
{\small that are UP or UC in the reduced network} & {\small 156} & {\small 117 } & {\small 182 } \\
\hline
\end{tabular}

\vspace{0.4cm} \noindent {\bf Table 1. Almost all globally
essential reactions in \ecoli, \yeast\ and \aureus\ are UP or
UC.\\}


\vspace{0.7cm}

\hspace{1cm}
\begin{tabular}{|c|c|c|c|}
\hline
\small Size of & \small Number of & \small Number of perfect & \small Breakup of explained clusters into\\
\small perfect clusters & \small perfect clusters & \small clusters explained & \small UP-UC clusters in the reduced network\\
\hline
\small 2 & \small 48 & \small 22 & \small 22 x (2) \\
\hline
\small 3 & \small 19 & \small 9 & \small 7 x (3) + 2 x (2) \\
\hline
\small 4 & \small 11 & \small 10 & \small 8 x (4) + 1 x (3) + 1 x (2) \\
\hline
\small 5 & \small 4 & \small 3 & \small 1 x (4) + 1 x (3+2) + 1 x (2+2) \\
\hline
\small 6 & \small 1 & \small 1 & \small 1 x (6) \\
\hline
\small 7 & \small 1 & \small 1 & \small 1 x (7) \\
\hline
\small 8 & \small 2 & \small 2 & \small 1 x (6+2) + 1 x (5+2) \\
\hline
\small 148 & \small 1 & \small 1 & \small (14+12+10+9+7+6+6+6+5+5+4+4+\\
& & & \small 4+4+4+3+3+3+2+2+2+2+2+2+2+2) \\
\hline
\end{tabular}

\vspace{0.4cm} \noindent {\bf Table 2. The size distribution of
perfect clusters in the \ecoli\ metabolic network and their
explanation in terms of UP-UC clusters.\\} The third column lists
the number of perfect clusters that are explained by UP-UC
clusters calculated using the reduced network. The fourth column
gives the breakup of the explained perfect clusters in terms of
UP-UC clusters of various sizes. E.g. in the second row the entry
7 x (3) + 2 x (2) implies that 7 UP-UC clusters of size 3 are
identical to 7 perfect clusters of size 3 and furthermore, two
UP-UC clusters of size 2 are subsets of two size 3 perfect
clusters. In the fourth row the term 1 x (3+2) means that one of
perfect clusters of size 5 contained two distinct UP-UC clusters
of sizes 3 and 2. There are 26 UP-UC clusters that are part of the
largest perfect cluster of 148 reactions accounting for 125
reactions in it. This largest perfect cluster is a subset of
reactions that are active for all input conditions and is located
near the output end of the metabolic network. \\

\newpage

\noindent
{\bf FIGURES\\}

\vspace{0.1cm}
\begin{center}
\fbox{
\begin{minipage}[b]{.50\textwidth}
\centering
{\bf (a)}
\includegraphics[width=2.7in]{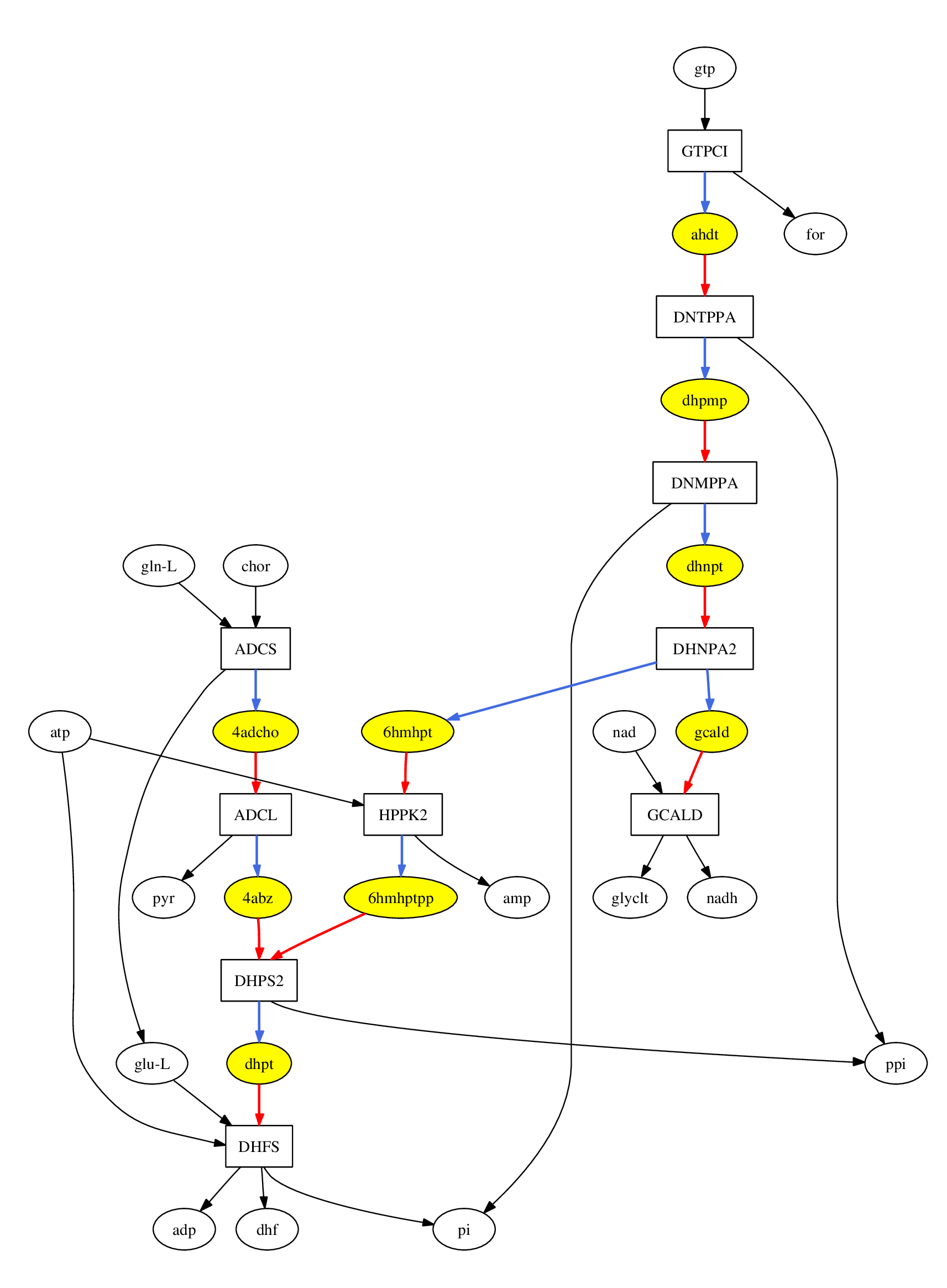}
\end{minipage}%
\begin{minipage}[b]{.50\textwidth}
\centering
{\bf (b)}
\includegraphics[width=2.3in]{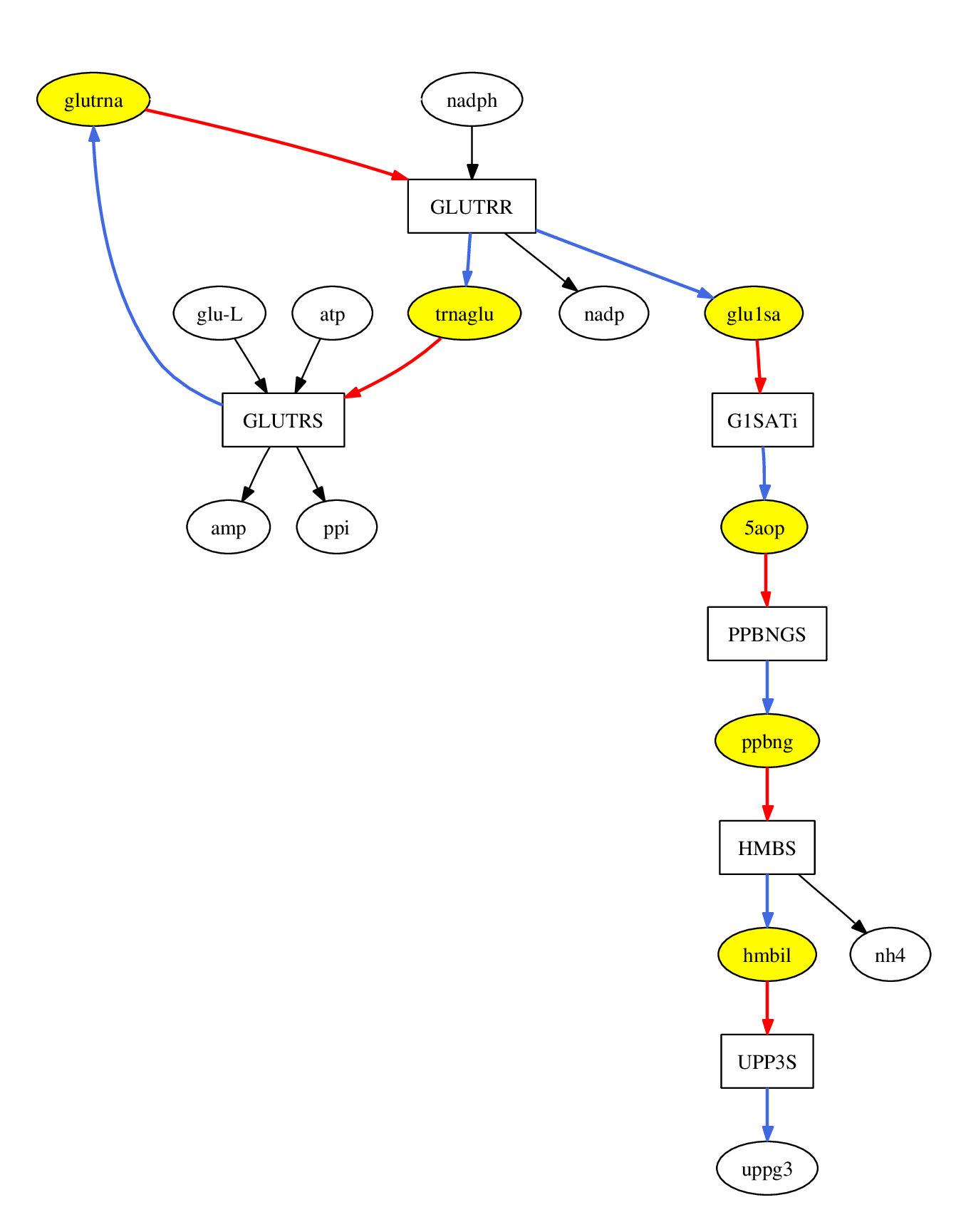}
\end{minipage}
}
\end{center}

\vspace{0.2cm}
\noindent
\small {{\bf Figure 1.} (a) UP-UC
metabolites in the \ecoli\ metabolic network forming a UP-UC
cluster of 10 reactions. (b) UP-UC metabolites in the \aureus\
metabolic network forming a UP-UC cluster of 6 reactions.
Rectangles represent reactions and ovals metabolites. Yellow ovals
represent UP-UC metabolites. Arrows to (from) metabolites
represent their production (consumption) in reactions. A blue
(red) link represents the production (consumption) of a UP (UC)
metabolite. Note that UP-UC clusters are not strictly linear
pathways. For example, in part (a) the reactions in the cluster
are not all in a single chain and in part (b) there is a cycle
inside the UP-UC cluster. Nevertheless fixing the flux of any one
reaction in a UP-UC cluster fixes the fluxes of all other
reactions in the cluster in any steady state, since the production
rate of every UP-UC metabolite must be the same as its consumption
rate. Hence, in part (a), fixing the flux of reaction GCALD fixes
the flux of reaction DHNPA2 (because of the intermediate UP-UC
metabolite gcald), which in turn fixes the fluxes of reactions
HPPK2 and DNMPPA, and so on. All reactions in part (a) and (b) are
globally essential in \ecoli\ and \aureus\ respectively. To reduce
clutter, nodes corresponding to $h$ (proton) and $h_2o$ have been
omitted. Abbreviation of metabolite and reaction names in part (a)
are as in \cite{RVSP2003} and in part (b) as in \cite{BP2005}.
The figures have been drawn using Graphviz software.\\
}


\newpage

\vspace{0.2cm}

\begin{center}
\includegraphics[width=2.7in]{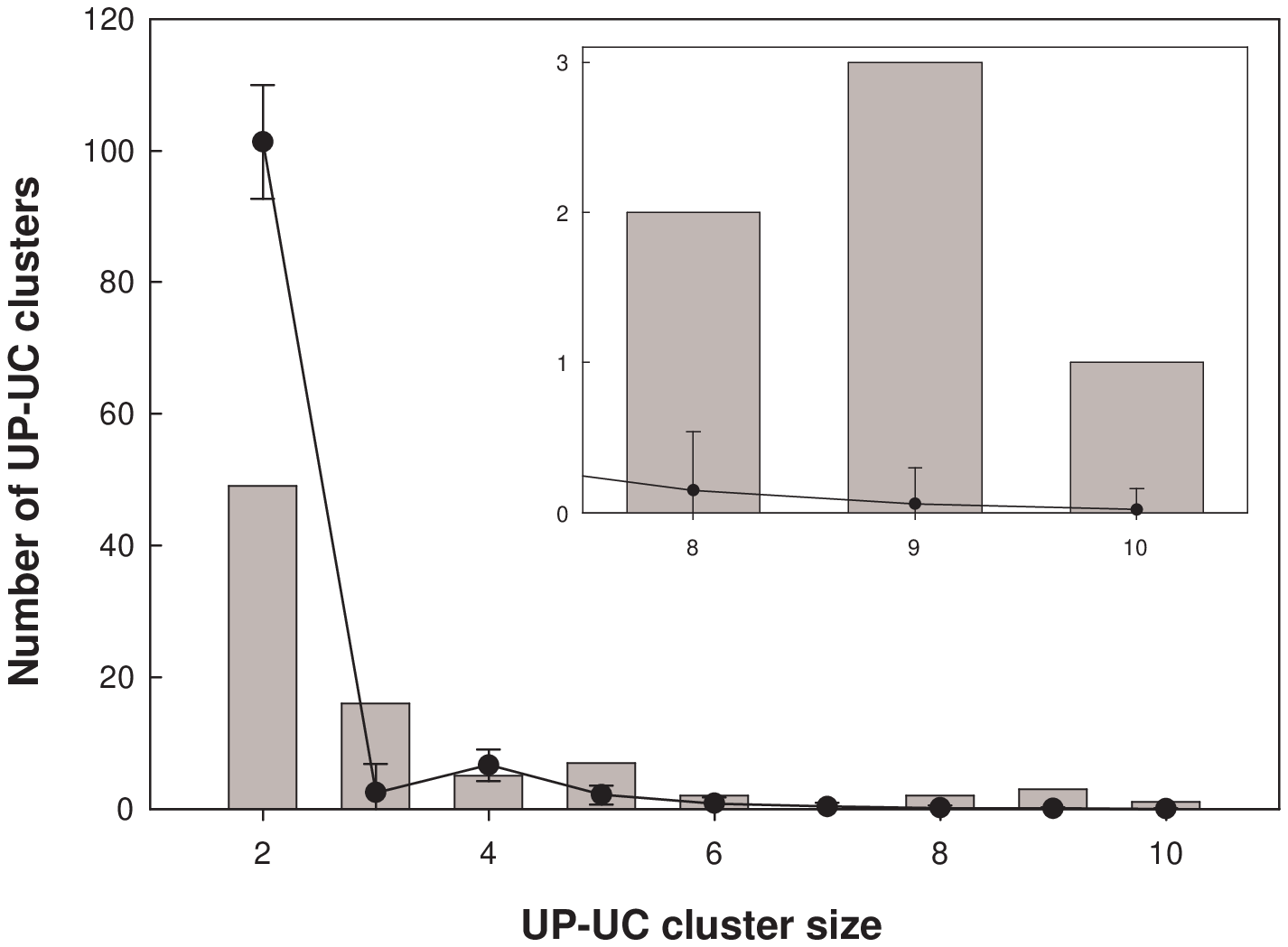}
\end{center}

\vspace{0.2cm} \noindent \small {{\bf Figure 2.} Frequency
histogram of UP-UC cluster sizes in the \ecoli\ metabolic network
(grey bars). Data is shown in Supplementary Table S5. The black
line is the frequency distribution for the randomized versions of
the network (averaged over 1000 realizations) that preserve the
in- and out-degree of all nodes. Error bars show one standard
deviation of the randomized ensemble. {\bf Inset:} Enlargement of
the graph for the larger sized clusters. In the real network,
larger UP-UC clusters (size $\geq 8$) occur much more often than
in the randomized version ($p < 0.001$). On the other hand,
smaller UP-UC clusters (size $\leq 3$) occur much less often than
in the randomized version ($p < 0.001$).\\
}


\end{document}